\documentstyle[prl,aps,multicol,psfig]{revtex}
\begin{document}
\title{\bf Quantum Computing by Two-Dimensional NMR
using Spin- and Transition-Selective Pulses\\}
\author{T.S. Mahesh $^\dagger$, Kavita Dorai $^\dagger$, Arvind $^\sharp$, and 
        Anil Kumar $^{\dagger,\ddagger}$\thanks{Author to whom correspondence should be addressed. 
        \newline    e-mail: \it{anilnmr@physics.iisc.ernet.in}}\\
	$^{\dagger}$ Department of Physics,  $^{\ddagger}$ Sophisticated Instruments Facility\\
	Indian Institute of Science, Bangalore 560012 India\\
	$^{\sharp}$  Department of Physics, Guru Nanak Dev University Amritsar 143005 India\\}
\maketitle

\begin{abstract}
Quantum computing using two-dimensional NMR has recently been described 
using scalar coupling evolution technique [J. Chem. Phys., \bf 109\normalfont, 
10603 (1998)].  In the present paper, we describe 
two-dimensional NMR quantum computing with the help of selective pulses.  A 
number of logic gates are implemented using two and three qubits with 
one extra observer spin.  Some many-in-one gates (or Portmanteau gates) 
are implemented.  Toffoli gate (or AND/NAND gate) and OR/NOR gates are 
implemented on three qubits.  Deutsch-Jozsa quantum algorithm for one 
and two qubits, using one extra work qubit, has also been implemented 
using selective pulses after creating a coherent superposition state, 
in the two-dimensional methodology.

\end{abstract}  

\begin{multicols}{2}
\section*{Introduction}
Quantum computing exploits the intrinsic quantum nature of 
physical systems and is therefore more powerful than classical computing 
for a certain class of problems.  While some problems like factorization gain 
exponential speed up (1), some others like database-search gain 
polynomial speed up (2).  A quantum computer works on two-level quantum systems 
known as `quantum bits' or `qubits'. The fact that qubits can exist in a coherent
superposition of basis states is exploited in a quantum computer.  However, 
retaining such a coherent superposition for long enough time is one of the 
major hurdles in quantum computation.  Since nuclear spins in liquids are 
efficiently isolated from the rest of their environment, the coherent
superposition lasts for several hundreds of milliseconds.  Therefore, high 
resolution NMR of weakly coupled nuclear spins in liquids is one of the 
potential candidates for realizing such a quantum computer.  Nuclear spins
with $\it{I}$=1/2 have generally been used since spins with $\it{I}$$>$1/2
relax much faster.    Quantum computing using one-dimensional NMR has been 
demonstrated by various workers (3-12).

	Recently, quantum computing has been demonstrated using 
two-dimensional NMR (13).  In this method, the states of the `computation'
spins (or input spins) are encoded by the transitions of an extra 
`observer' spin (Fig.1).
 The observer spin is first allowed to 
evolve for a time t$_{1}$ during which, the input qubits remain in their initial 
state (Fig.2). After the frequency labelling period t$_{1}$, the computation is 
performed on the input qubits.  The observer spin is again allowed to evolve 
for a time t$_{2}$ and detected. The experiment starts from a mixed 
input state, the computation is performed simultaneously on all the 
input states, and one gets a mixed output state. One can also create a 
superposition of input qubits in the beginning of the experiment by applying 
a $\pi$/2 pulse on the input qubits and subsequently killing the coherence 
by a gradient pulse (13).  Various steps in two-dimensional NMR namely, 
preparation, evolution, mixing and detection have a close correspondence with 
the steps in quantum computing namely, creation of initial states, labeling 
of the initial states, computation and reading of output states, respectively 
(13) (Fig.2). In this respect, two-dimensional NMR appears to be the `method of 
choice' for quantum computing. 

An advantage of the two-dimensional method is that it resolves all the input
and the output states and correlates them.  This correlation between input 
and output states in the two-dimensional experiment makes the result of the 
computation graphic.  For example,  a SWAP gate which exchanges the 
states of two qubits, can be implemented in NMR by selectively interchanging 
populations of zero-quantum levels.  Since the one-dimensional NMR spectrum 
of homonuclear spins after the execution of SWAP gate is indistinguishable 
from that of an equilibrium spectrum, the one-dimensional SWAP gate was 
earlier demonstrated after the creation of a non-equilibrium state (12, 14).  
However, as will be shown here (Fig.3), and has been demonstrated earlier 
(13), the two-dimensional method starts from the equilibrium state and 
yields a spectrum characteristic of the SWAP gate.

Two-dimensional quantum computation has been carried out so far by using scalar 
coupling evolution technique (13). This paper describes a two dimensional 
NMR implementation of several two and three qubit gates with one extra 
observer spin, using spin- and transition-selective pulses.  The first 
implementation of the Deutsch-Jozsa quantum algorithm using 
two-dimensional NMR is also demonstrated here.  One of the advantages of the 
selective pulse method is that it makes the computation simple and straightforward.

\section*{Experimental}
\subsection*{A. Logic Gates}
A weakly coupled 3-spin system (I$_{0}$I$_{1}$I$_{2}$) can be viewed as a 
combination of two 2-spin subsystems with the observer spin (I$_{0}$) being 
in state 0 or 1 (Fig.1).  
\begin{figure}
\begin{center}
\hspace{.5cm}
\psfig{file=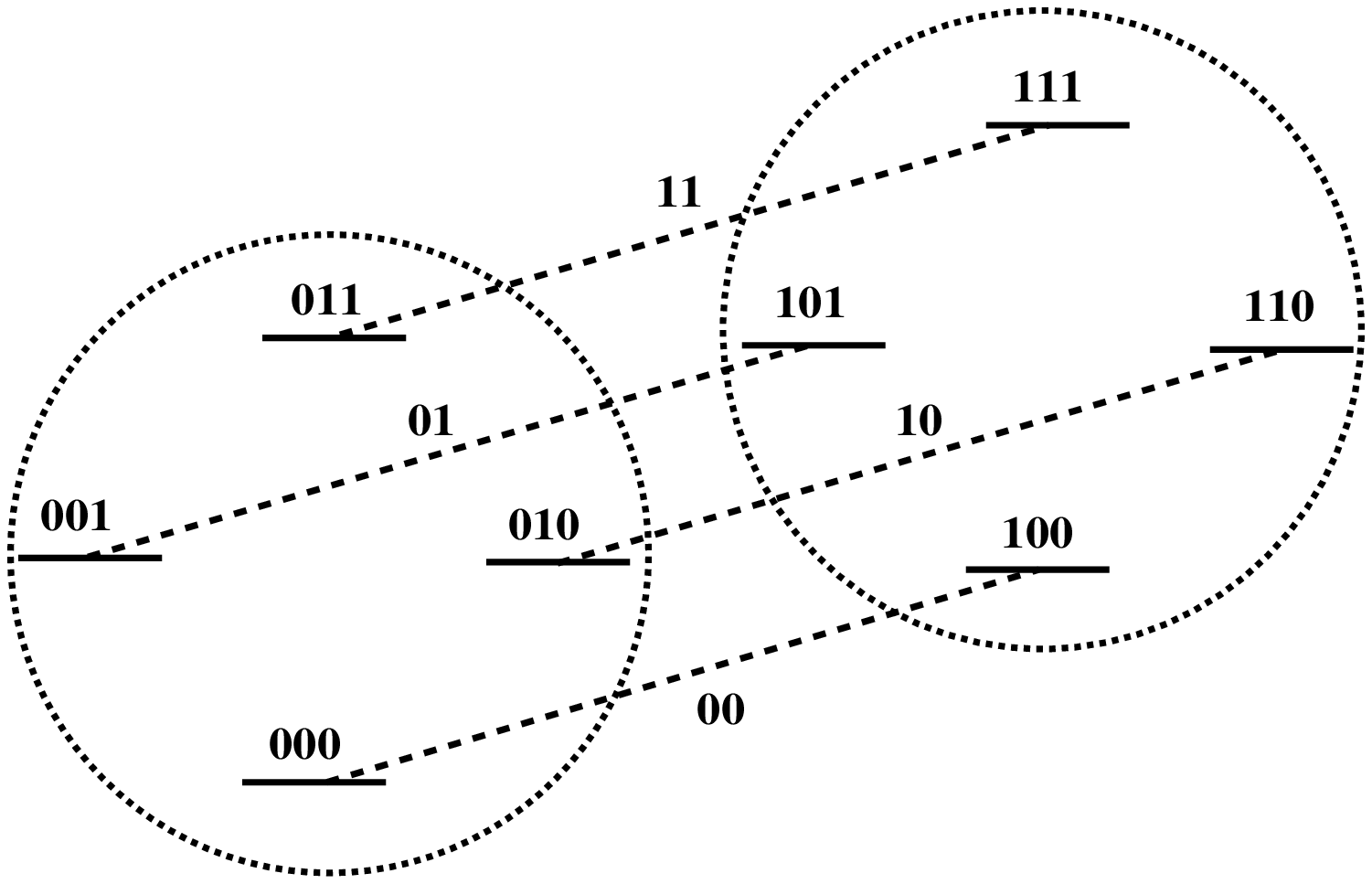,width=8.5cm}
\end{center}
\bf{Figure 1.} \normalfont \small Energy level diagram of a 3-spin system 
(I$_{0}$I$_{1}$I$_{2}$) showing two 2-spin subsystems corresponding to 
states 0 and 1 of observer spin I$_{0}$.  The transitions of I$_{0}$
are labelled by the states of input qubits (I$_{1}$I$_{2}$).\normalsize
\end{figure}
There are four observer spin transitions labelled 
as 11, 10, 01 and 00, which correspond to the states of the input qubits 
(I$_{1}$I$_{2}$) in these transitions.  During the computation (mixing) 
period, various transitions of the input qubits are inverted yielding various 
gates as listed in Table 1. The results of 24 one-to-one 
reversible gates for the 3-spin system having one observer and two input qubits 
are shown in Fig.3. The NOP gate is implemented by doing No OPeration during 
the computation period.  In this gate, each input state corresponds to same 
output state after the computation.  Various NOT gates are implemented by 
inverting one or both of input qubits (I$_{1}$,I$_{2}$), using spin-selective 
$\pi$ pulses.
       XOR (or control-NOT) and XNOR gates are implemented by inverting 
two similar transitions of the same input spin, one in each subsystem.  We label XOR 
and XNOR gates as XOR1, XNOR1 and XOR2, XNOR2 depending on whether the result 
of the operation is stored on the spin I$_{1}$ or I$_{2}$ respectively.  For 
example, XOR1 gate is implemented by inverting transitions 001-011 and 101-111 
of I$_{1}$ and the result of the gate is stored on I$_{1}$ (Table 1, Fig.3). A 
SWAP gate can be implemented by selectively interchanging the populations of 
zero quantum levels of each subsystem (Table 1, Fig.3).  This can be achieved 
by the use of a cascade of three transition-selective, non-commuting $\pi$ 
pulses on regressively connected transitions during the computation period (12, 14).
It may be noted that while all the gates of ref.(13) are reproduced here, with 
several new gates added, the methodology used for the computation is different 
except for the NOP and NOT gates.

Similar to classical Boolean expressions, quantum gates can be reduced to a 
sequence of simple unitary operations.  Although a few basic gates such as NOT, 
AND, OR, XOR are sufficient to carryout a given computation, other gates are 
also useful since they may reduce the number of pulses.  Therefore `portmanteau 
gates' which perform more than one operation have been implemented in 
one-dimensional NMR using selective pulses (12).

\begin{figure}
\begin{center}
\hspace{.5cm}
\psfig{file=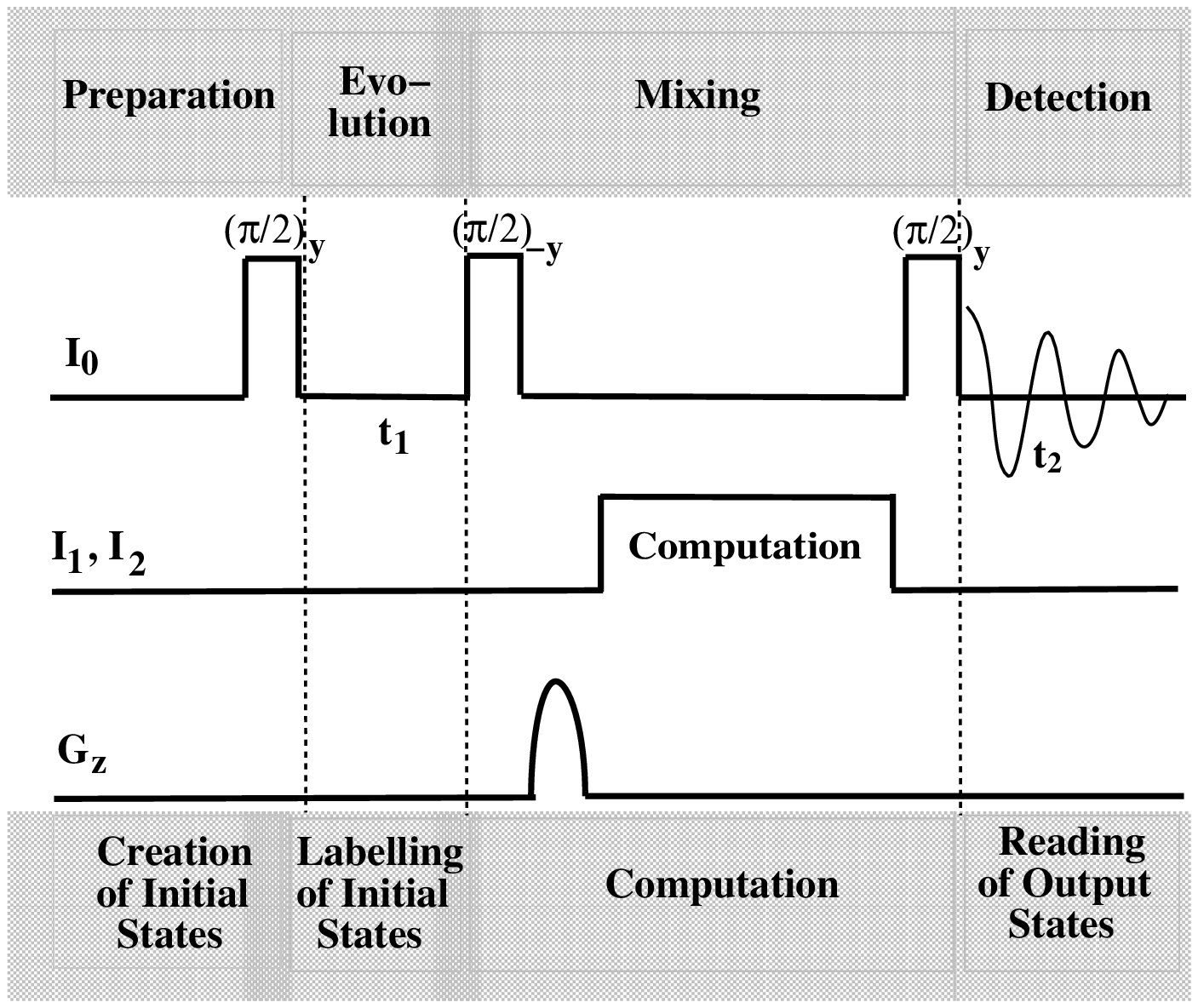,width=8.5cm}
\end{center}
\bf{Figure 2.} \normalfont \small Pulse scheme for the two-dimensional NMR implementation of 
quantum computing.  The close correspondence between two-dimensional
NMR and quantum computing is shown. I$_{0}$ is the observer spin and 
I$_{1}$, I$_{2}$ are the input qubits.  The gradient pulse G$_{z}$ eliminates the unwanted
transverse magnetization before the computation. \normalsize
\end{figure}

\end{multicols}
\begin{figure}
\begin{center}
\hspace{.5cm}
\psfig{file=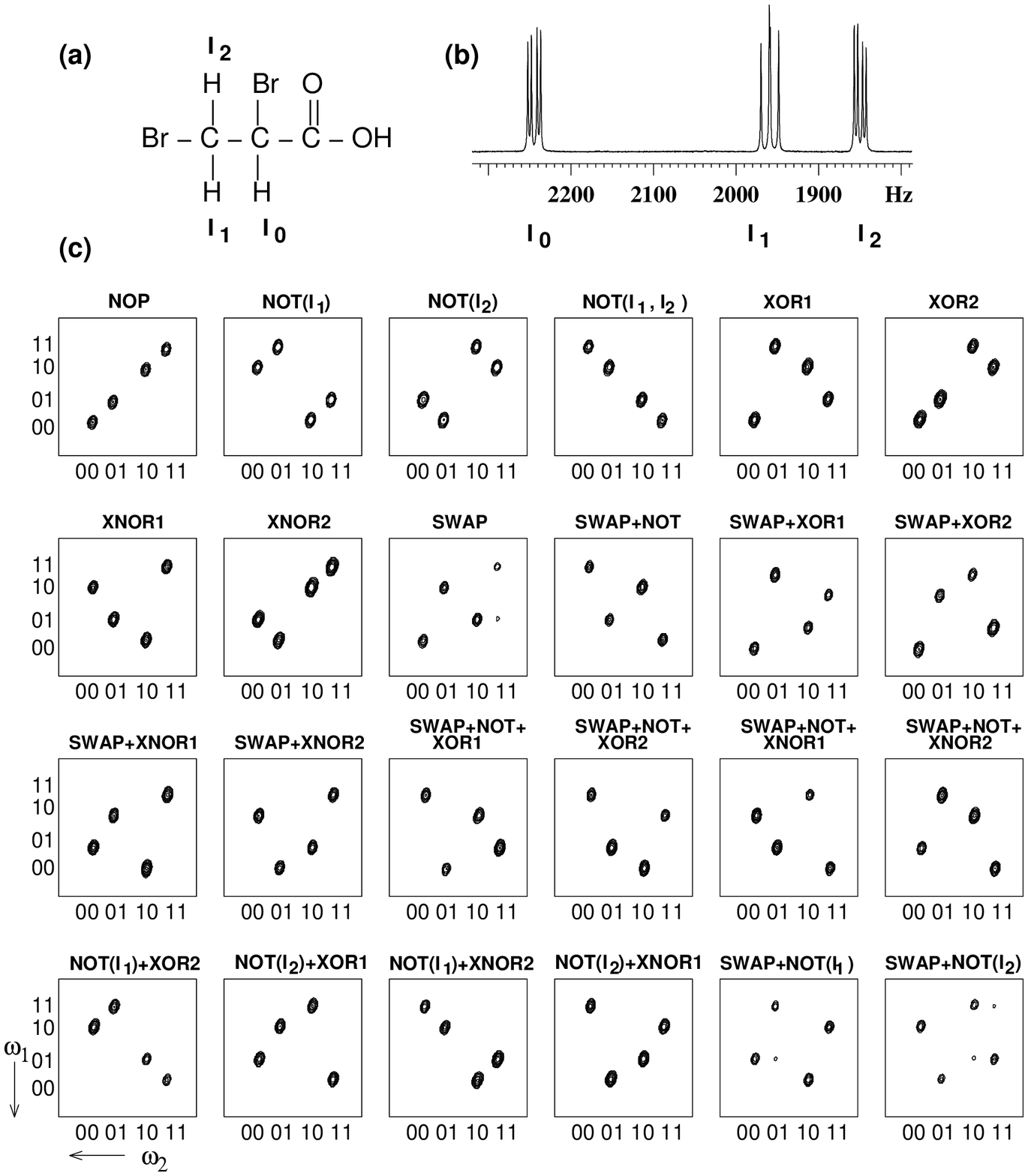,height=19cm}
\end{center}
\bf{Figure 3.} \normalfont \small $^{1}$H NMR spectrum (b) of 2,3-dibromo-propionic 
acid (a) in CDCl$_{3}$ recorded on a Bruker DRX-500 spectrometer at 300K. 
(c) shows observer spin (I$_{0}$) spectra corresponding to various 
gates implemented using spin- and transition-selective pulses.  Pulse scheme 
used is shown in Fig.2 and various transitions of input qubits (I$_{1}$, I$_{2}$)
inverted during the computation are listed in Table 1.  The 
spin-selective pulses were 10ms long and the transition-selective pulses 
were 100-300ms long.  The phase of the computation pulses were cycled through 
($x, -x$) to
suppress  the distortions due to pulse imperfections.   All experiments were 
carried out in the time domain with 256 t$_{1}$ values and 256 complex data 
points along t$_{2}$ and with 2 scans for each t$_{1}$ point.  Zero filling to
512$\times$512 complex data points was done prior to the 2D Fourier
transformation.  All plots are shown in magnitude mode. \normalsize
\end{figure}
\newpage
\begin{center}
\bf{TABLE 1} \\
\bf Various logic gates with Boolean algebra, truth table and operations
performed
\footnotesize
\renewcommand{\tabcolsep}{.5cm}
\begin{tabular}{|p{2cm}|p{.2cm}|p{.2cm}|p{1.5cm}||p{2cm}|p{.2cm}|p{.2cm}|p{1.5cm}|}
\hline

 {\bf $\;\;\;\;\;\;\;\;\;\;\;$ GATE$^{a}$}&{$\;\;\;$ \footnotesize IN}
&{$\;$ \footnotesize OUT}&{\bf Transitions Inverted$^{b}$}		
\addtocounter{footnote}{-1}
&{\bf $\;\;\;\;\;\;\;\;\;\;\;$ GATE$^{a}$}&{$\;\;\;$ \footnotesize IN}
&{$\;$ \footnotesize OUT}&{\bf Transitions Inverted$^{b}$} \\
\hline

 {\bf NOP $\;\;\;\;\;\;$ $|s,t\rangle$$\rightarrow$$|s,t\rangle$}		
&{11 10 01 00}	&{11 10 01 00}	&{No OPeration}
&{\bf NOT(I$_{1}$) $\;\;\;\;\;$ $|s,t\rangle$$\rightarrow$$|\overline{s},t\rangle$}	
&{11 10 01 00}	&{01 00 11 10}	&{All transitions of I$_{1}$} \\      
\hline

 {\bf NOT(I$_{2}$) $\;\;\;\;\;$ $|s,t\rangle$$\rightarrow$$|s,\overline{t}\rangle$}		
&{11 10 01 00}	&{10 11 00 01}	&{All transitions of I$_{2}$}
&{\bf NOT(I$_{1}$,I$_{2}$) $\;\;\;$ $|s,t\rangle$$\rightarrow$$|\overline{s,t}\rangle$}	
&{11 10 01 00}	&{00 01 10 11}	&{All transitions of  I$_{1}$ and I$_{2}$} \\      
\hline

 {\bf XOR1 $\;\;\;\;\;\;$ $|s,t\rangle$$\rightarrow$$|s$$\oplus$$t,t\rangle$} 
&{11 10 01 00}	&{01 10 11 00}	&{111 $\leftrightarrow$ 101 011 $\leftrightarrow$ 001}
&{\bf XOR2 $\;\;\;\;\;\;$ $|s,t\rangle$$\rightarrow$$|s,s$$\oplus$$t\rangle$} 
&{11 10 01 00}	&{10 11 01 00}	&{111 $\leftrightarrow$ 110 011 $\leftrightarrow$ 010} \\
\hline

 {\bf XNOR1 $\;\;\;\;\;\;$ $|s,t\rangle$$\rightarrow$$|\overline{s\oplus t},t\rangle$} 
&{11 10 01 00}	&{11 00 01 10}	&{100 $\leftrightarrow$ 110 000 $\leftrightarrow$ 010}
&{\bf XNOR2 $\;\;\;\;\;\;$ $|s,t\rangle$$\rightarrow$$|s,\overline{s\oplus t}\rangle$} 
&{11 10 01 00}	&{11 10 00 01}	&{101 $\leftrightarrow$ 100 001 $\leftrightarrow$ 000} \\
\hline

 {\bf SWAP $\;\;\;\;\;\;$ $|s,t\rangle$$\rightarrow$$|t,s\rangle$} 
&{11 10 01 00}	&{11 01 10 00}	&{\bf Invert$\;$ZQ: \normalfont
				  110 $\leftrightarrow$ 111 
				  010 $\leftrightarrow$ 011
				  101 $\leftrightarrow$ 111
				  001 $\leftrightarrow$ 011
				  110 $\leftrightarrow$ 111
				  010 $\leftrightarrow$ 011}
&{\bf SWAP+NOT  $|s,t\rangle$$\rightarrow$$\overline{|t,s\rangle}$} 
&{11 10 01 00}	&{00 10 01 11}	&{\bf Invert$\;$DQ: \normalfont
				  110 $\leftrightarrow$ 111 
				  010 $\leftrightarrow$ 011
				  100 $\leftrightarrow$ 110
				  000 $\leftrightarrow$ 010
				  110 $\leftrightarrow$ 111
				  010 $\leftrightarrow$ 011} \\
\hline

 {\bf SWAP+XOR1 $|s,t\rangle$$\rightarrow$$|s$$\oplus$$t,s\rangle$} 
&{11 10 01 00}	&{01 11 10 00}	&{101 $\leftrightarrow$ 111
				  001 $\leftrightarrow$ 011 
				  110 $\leftrightarrow$ 111
				  010 $\leftrightarrow$ 011}
&{\bf SWAP+XOR2 $|s,t\rangle$$\rightarrow$$|t,s$$\oplus$$t\rangle$} 
&{11 10 01 00}	&{10 01 11 00}	&{110 $\leftrightarrow$ 111
				  010 $\leftrightarrow$ 011 
				  101 $\leftrightarrow$ 111
				  001 $\leftrightarrow$ 011} \\
\hline

 {\bf SWAP+ XNOR1 $|s,t\rangle$$\rightarrow$$|\overline{s\oplus t},s\rangle$} 
&{11 10 01 00}	&{11 01 00 10}	&{100 $\leftrightarrow$ 110
				  000 $\leftrightarrow$ 010 
				  100 $\leftrightarrow$ 101
				  000 $\leftrightarrow$ 001}
&{\bf SWAP+ XNOR2 $|s,t\rangle$$\rightarrow$$|t,\overline{s\oplus t}\rangle$} 
&{11 10 01 00}	&{11 00 10 01}	&{100 $\leftrightarrow$ 101
				  000 $\leftrightarrow$ 001
				  100 $\leftrightarrow$ 110
				  000 $\leftrightarrow$ 010} \\
\hline

 {\bf SWAP+NOT+ XOR1 $|s,t\rangle$$\rightarrow$$|\overline{s}$$\oplus$$\overline{t},\overline{s}\rangle$} 
&{11 10 01 00}	&{00 10 11 01}	&{101 $\leftrightarrow$ 111
				  001 $\leftrightarrow$ 011 
				  100 $\leftrightarrow$ 101
				  000 $\leftrightarrow$ 001}
&{\bf SWAP+NOT+ XOR2 $|s,t\rangle$$\rightarrow$$|\overline{t},\overline{s}$$\oplus$$\overline{t}\rangle$} 
&{11 10 01 00}	&{00 11 01 10}	&{110 $\leftrightarrow$ 111
				  010 $\leftrightarrow$ 011 
				  100 $\leftrightarrow$ 110
				  000 $\leftrightarrow$ 010} \\
\hline

 {\bf SWAP+NOT+ XNOR1 $|s,t\rangle$$\rightarrow$$|\overline{\overline{s}\oplus \overline{t}},\overline{s}\rangle$} 
&{11 10 01 00}	&{10 00 01 11}	&{100 $\leftrightarrow$ 110
				  000 $\leftrightarrow$ 010 
				  110 $\leftrightarrow$ 111
				  010 $\leftrightarrow$ 011}
&{\bf SWAP+NOT+ XNOR2 $|s,t\rangle$$\rightarrow$$|\overline{t},\overline{\overline{s}\oplus \overline{t}}\rangle$} 
&{11 10 01 00}	&{01 10 00 11}	&{100 $\leftrightarrow$ 101
				  000 $\leftrightarrow$ 001
				  101 $\leftrightarrow$ 111
				  001 $\leftrightarrow$ 011} \\
\hline

 {\bf NOT(I$_{1}$)+ XOR2 $|s,t\rangle$$\rightarrow$$|\overline{s},\overline{s}$$\oplus$$t\rangle$} 
&{11 10 01 00}	&{01 00 10 11}	&{All I$_{1}$ transitions and $\;\;$ 111 $\leftrightarrow$ 110 011 $\leftrightarrow$ 010}
&{\bf NOT(I$_{2}$)+ XOR1 $|s,t\rangle$$\rightarrow$$|s$$\oplus$$\overline{t},\overline{t}\rangle$} 
&{11 10 01 00}	&{10 01 00 11}	&{All I$_{2}$ transitions and $\;\;$ 111 $\leftrightarrow$ 101 011 $\leftrightarrow$ 001} \\
\hline

 {\bf NOT(I$_{1}$)+ XNOR2 $|s,t\rangle$$\rightarrow$$|\overline{s},\overline{\overline{s}\oplus t}\rangle$} 
&{11 10 01 00}	&{00 01 11 10}	&{All I$_{1}$ transitions and $\;\;$ 101 $\leftrightarrow$ 100 001 $\leftrightarrow$ 000} 
&{\bf NOT(I$_{2}$)+ XNOR1 $|s,t\rangle$$\rightarrow$$|\overline{s\oplus \overline{t}},\overline{t}\rangle$} 
&{11 10 01 00}	&{00 11 10 01}	&{All I$_{2}$ transitions and $\;\;$ 100 $\leftrightarrow$ 110 000 $\leftrightarrow$
010} \\
\hline

 {\bf SWAP+ NOT(I$_{1}$) $|s,t\rangle$$\rightarrow$$|\overline{t},s\rangle$} 
&{11 10 01 00}	&{01 11 00 10}	&{Invert ZQ and all transitions of I$_{1}$}
&{\bf SWAP+ NOT(I$_{2}$) $|s,t\rangle$$\rightarrow$$|t,\overline{s}\rangle$} 
&{11 10 01 00}	&{10 00 11 01}	&{Invert ZQ and all transitions of I$_{2}$} \\
\hline
\end{tabular}
\end{center}

\footnotesize
\indent
$^{a}$ $|s,t\rangle$ represents the state of input qubits (I$_{1}$,I$_{2}$). \\ 
\indent
$^{b}$ Order of transitions are important as all transitions may not commute.
\normalsize

\begin{multicols}{2}
\noindent
Similar gates have been 
implemented here using two-dimensional NMR (Table 1, Fig.3). Interchanging 
populations of double quantum levels using a cascade of non-commuting $\pi$ 
pulses on progressively connected transitions leads to SWAP+NOT gate (Table 1, 
Fig.3).  SWAP+XOR and SWAP+XNOR gates have been implemented by inverting two 
pairs of regressively connected transitions in each case (Table 1, Fig.3).  
Similarly, SWAP+NOT+XOR and SWAP+NOT +XNOR gates have been implemented by 
inverting two pairs of progressively connected transitions (Table 1, Fig.3).  
The last six gates in Fig.3 are direct combinations of two gates.  It may be 
noted that the 24 gates shown in Fig.3 form a complete set of 2-qubit one-to-one
mappings.

\begin{figure}
\begin{center}
\hspace{-2cm}
\psfig{file=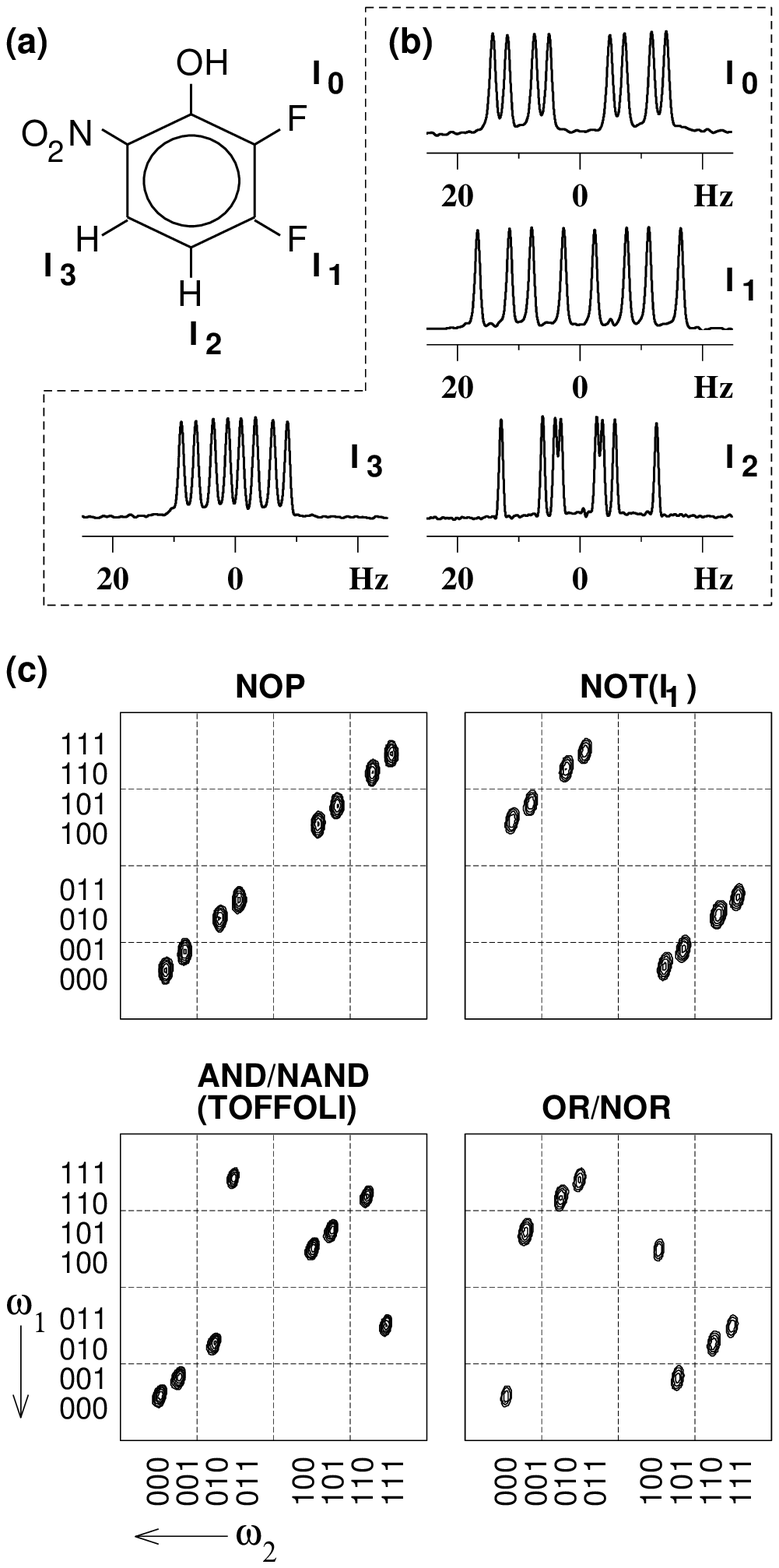,width=9cm}
\end{center}
\vspace{-1cm}
\bf{Figure 4.} \normalfont \small $^{19}$F and $^{1}$H NMR spectra (b) of 
2,3-difluro-6-nitrophenol (a) in CDCl$_{3}$ (with one drop of D$_{2}$O
to induce the exchange of the hydroxy proton and hence to suppress
its coupling to flourine nuclei) recorded on a Bruker DRX-500 spectrometer 
at 300K. (c) shows $^{19}$F spectra of observer
spin I$_{0}$ corresponding to various gates. Pulse scheme 
used is shown in Fig.2 and various transitions of control qubit I$_{1}$
inverted during the computation period are described in the text.
The spin-selective pulses were 1ms long and the transition-selective pulses
were 200ms long.
The phase of the computation pulses were cycled through ($x, -x$) to
suppress distortions due to pulse imperfections.   All experiments were 
carried out in the time domain with 1024 t$_{1}$ values and 256 complex data 
points along t$_{2}$ and with 2 scans for each t$_{1}$ point.  Zero filling to
1024$\times$1024 complex data points was done prior to the 2D Fourier
transformation.  All plots are shown in magnitude mode. \normalsize
\end{figure}

   Fig.4 shows several 3-qubit gates implemented on a 4-spin system, using 
selective pulses.  Once again, No OPeration during computation period yields 
NOP gate and inverting spin I$_{1}$ yields the NOT(I$_{1}$) gate.  The 
more interesting ones are - Toffoli gate (or AND/NAND gate) and OR/NOR gate.
The operations of Toffoli and OR/NOR gates (13,15) are respectively,
\begin{equation}
|s,t,u\rangle\rightarrow|s \oplus (t \wedge u),t,u\rangle
\end{equation}
\begin{center}
and
\end{center}
\begin{equation}
|s,t,u\rangle\rightarrow|s \oplus (t \vee u),t,u\rangle,
\end{equation}
where $\oplus$ $\equiv$ addition modulo 2, $\wedge$ $\equiv$ AND, 
$\vee$ $\equiv$ OR and $s$, $t$, $u$ are the states of the control spin 
I$_{1}$ and the input spins I$_{2}$ and I$_{3}$.
The Toffoli gate is a universal gate for reversible computation.  This gate 
can be implemented by inverting the control spin transitions 011 and 111, 
by using a pair of transition-selective $\pi$ pulses.  Inverting all 
transitions of the control spin except the transitions 010 and 110 leads to 
OR/NOR gate (Fig.4).

\subsection*{B. Deutsch-Jozsa algorithm}
The Deutsch-Jozsa (DJ) algorithm demonstrates the power of quantum 
computing over classical computing (16).  In this algorithm, we consider
functions from N-bit domain space to 1-bit range space.
A function is called constant 
if it gives the same output for any input, and is called balanced if it gives one 
output for half the number of inputs and another for the remaining half. 
Given an N-bit binary function which is either constant or balanced, 
classically $2^{(N-1)}$+1 operations are required to determine whether the 
function is constant or balanced, whereas DJ algorithm requires only a single 
evaluation.  The DJ algorithm has been implemented using one-dimensional NMR by 
several research groups (10-13).

In the Cleve version of the DJ algorithm (17), a binary function $\it{f}$ is encoded 
in a unitary transformation by the propagator U$_f$ by including an extra work 
qubit such that 
\begin{equation}
|r \rangle |s \rangle{\stackrel{U_{f}}{\rightarrow}} 
|r \oplus f(s) \rangle |s \rangle, 
\end{equation}
where $|r \rangle$ and $|s \rangle$ are respectively the states of work qubit (I$_{0}$)
and input qubit (I$_{1}$).  The four possible functions for the 
single-bit DJ algorithm are listed in Table 2.  

\begin{center}
\bf TABLE 2\\
The four possible binary functions \\
($f_{1}$-$f_{4}$) for 1-bit DJ algorithm\\
\normalfont
\vspace{.5cm}
\renewcommand{\tabcolsep}{.5cm}
\begin{tabular}{|c|c|c|c|c|}
\hline
& 
\multicolumn{2}{c|}{CONST.} & 
\multicolumn{2}{c|}{BAL.}     \\ \cline{2-5}  
{\it{s}} & {\it{f$_{1}$}} & {\it{f$_{2}$}} & {\it{f$_{3}$}} & {\it{f$_{4}$}}
\\ \hline
0 & 0 & 1 & 0 & 1 \\
1 & 0 & 1 & 1 & 0 \\
\hline
\end{tabular}
\end{center}

\vspace{2cm}
\begin{figure}
\begin{center}
\vspace{2cm}
\hspace{.5cm}
\psfig{file=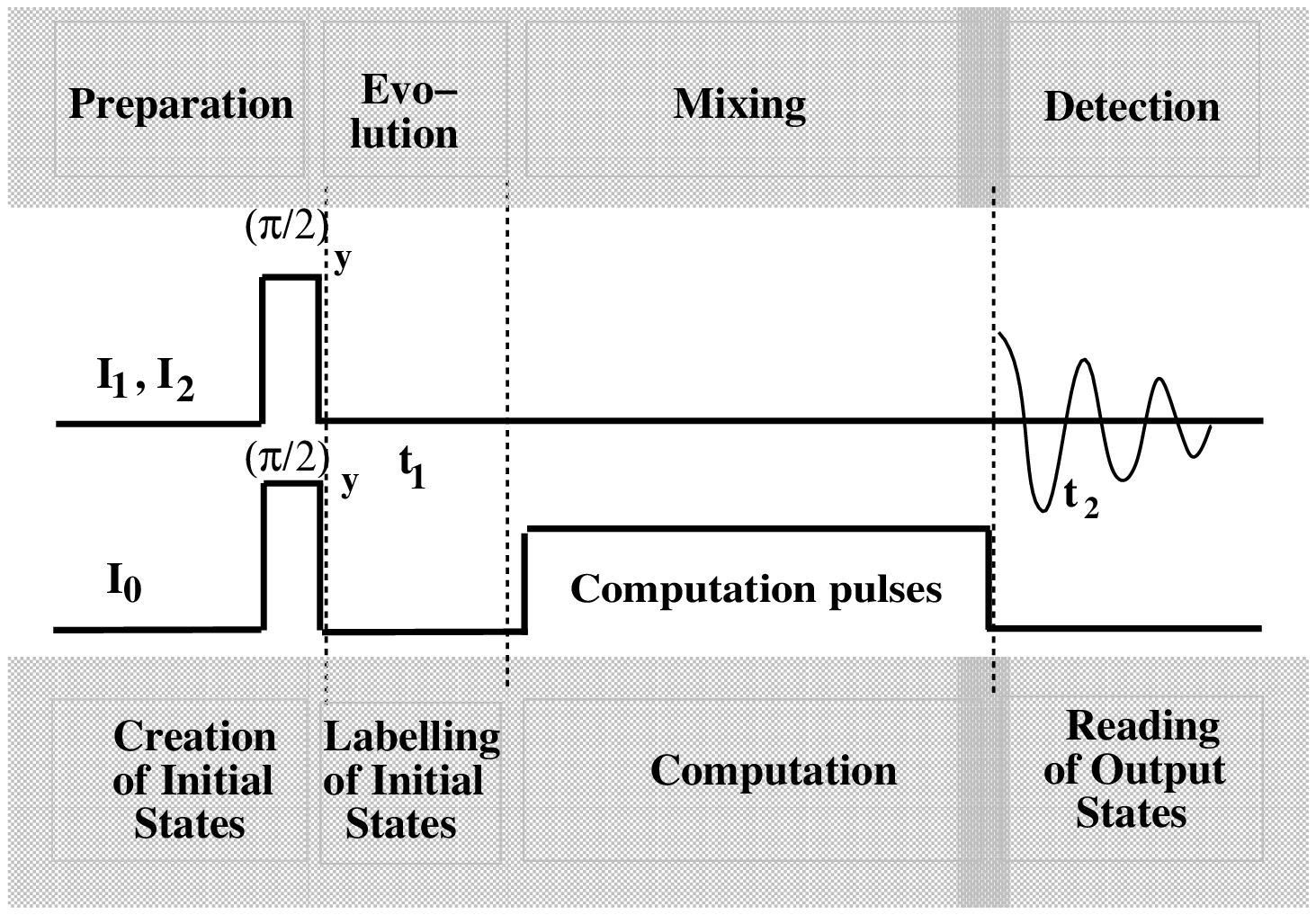,width=9cm}
\end{center}
\bf{Figure 5.} \normalfont \small Pulse scheme for the two-dimensional NMR implementation of 
DJ algorithm.   I$_{0}$ is the work qubit and I$_{1}$,I$_{2}$ are the input qubits. \\ \normalsize
\end{figure}

This algorithm has been implemented
on the thermal equilibrium state and does not require the creation of
a pure initial state (10,12).  The
two-dimensional pulse scheme used for implementing DJ algorithm is shown in Fig.5.
The experiment begins with
both qubits in a superposition of states, achieved by a non-selective
$(\pi/2)$$_{y}$ pulse.  This is followed by an evolution period t$_{1}$,
propagator U$_{f}$ and detection period t$_{2}$.
The transformations
corresponding to $\it{f}$$_{1}$ and $\it{f}$$_{2}$ are respectively, a unity
operation and a spin selective $\pi$$_{x}$ pulse on the work qubit.  The
transformations corresponding to $f_{3}$ and $f_{4}$ are implemented by
transition-selective $\pi$$_{x}$ pulses respectively on the 10-11 and 00-01
transitions of the work qubit (12).  The results of the algorithm for all the 4
functions are shown in Fig.6.
The constant or balanced nature of the function
is identified by the presence or absence of signal from the input qubit (I$_{1}$)
(Fig.6).  The expected signals are also shown schematically in Fig.6.  For two of the
functions $f_{2}$ and $f_{4}$, the expected signals are calculated in Table 3.
The spin-states ($|r \rangle |s \rangle$) before computation can be paired
(connected by a curved line) in such a way that each pair represents a transition
of the input qubit (I$_{1}$).  Each transition of the input qubit is labelled by
the state of the remaining spin i.e., work qubit.  Corresponding output pairs of spin-states 
after computation can be calculated by using expression [3].
From each output pair, the corresponding transition of the input spin after computation
can be identified.  If two spins have flipped in the output pair, then
the transition becomes non-observable and will be labelled as zero quantum (ZQ) or
double quantum (DQ).

\begin{figure}
\begin{center}
\psfig{file=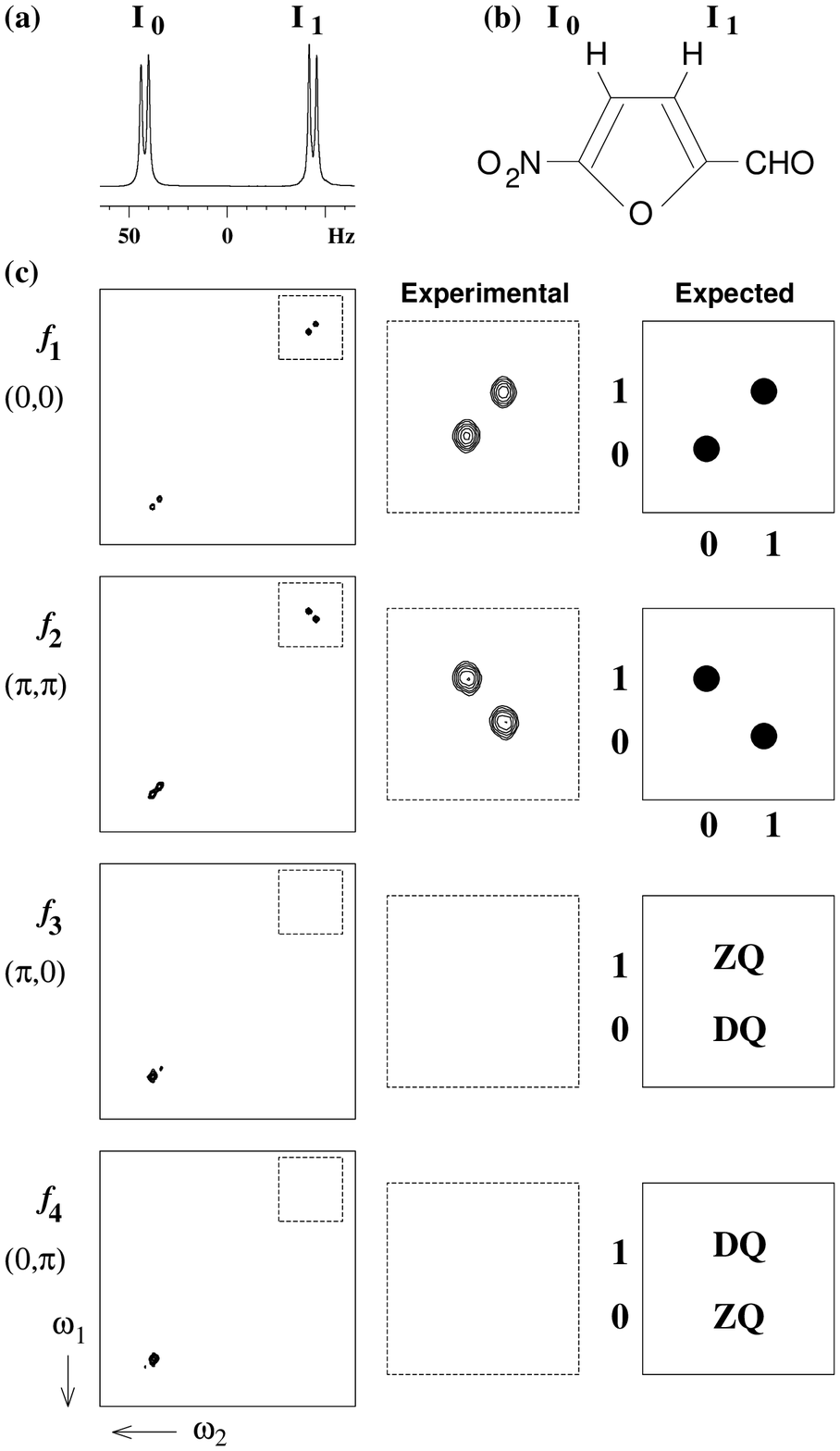,width=8.5cm}
\vspace{-1cm}
\end{center}
\bf{Figure 6.} \normalfont \small $^{1}$H NMR spectrum (a) of 5-nitro furaldehyde (b) in 
C$_{6}$D$_{6}$ on a Bruker DRX-500 spectrometer at 300K.  The results of DJ 
algorithm for various functions $f_{1}$-$f_{4}$ in Table 2 are shown in (c).  
I$_{0}$ is the work qubit and I$_{1}$ is the input qubit.  Expansions of the
I$_{1}$ parts of the spectra are shown in dotted boxes. The expected pattern 
is also shown in each case.  The expected patterns for $f_{2}$ and $f_{4}$
are described in Table 3. The spin-selective pulses 
were 10ms long and the transition-selective pulses were 100ms long.  The phase of the 
computation pulses were cycled through ($x, -x$) to suppress  the distortions due to 
pulse imperfections.   All experiments were carried out in the time domain with 
256 t$_{1}$ values and 512 complex data points along t$_{2}$ and with 2 scans for 
each t$_{1}$ point.  Zero filling to 512$\times$512 complex data points was done 
prior to the 2D Fourier transformation.  All plots are shown in magnitude mode. \\ \normalsize
\end{figure}

\newpage

\begin{center}
\bf TABLE 3\\
The input-output correlations for the functions $f_{2}$ and $f_{4}$ 
of 1-bit DJ algorithm.\\
\end{center}
\begin{figure}
\hspace{.2cm}
\psfig{file=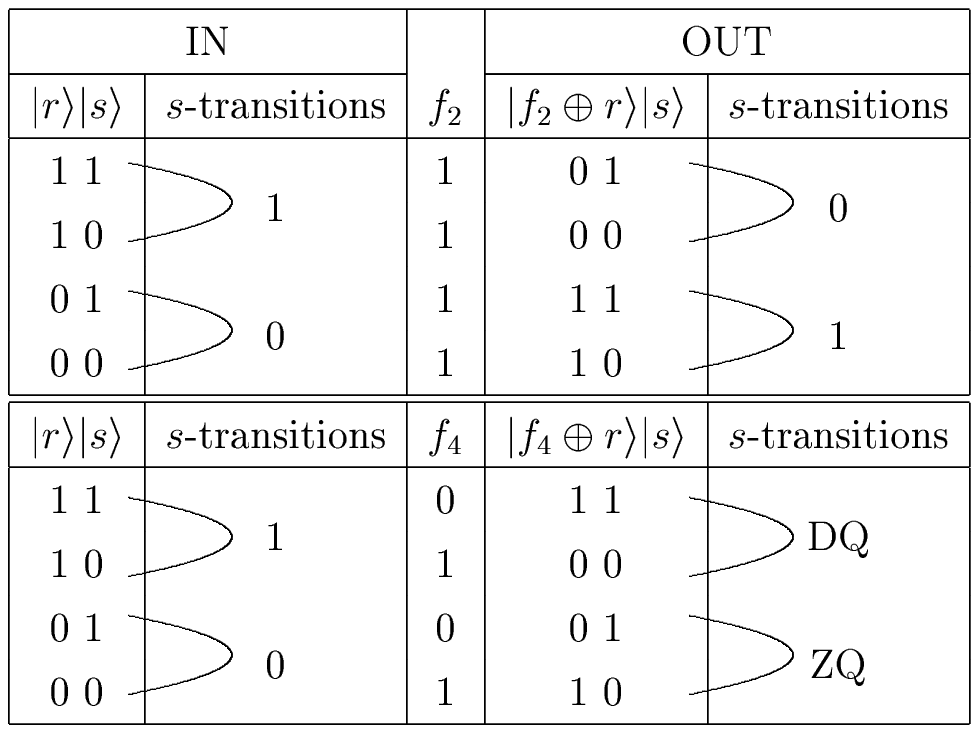,width=8.5cm,clip=,bbllx=4.9cm,bburx=15cm,bblly=16cm,bbury=25cm}
\end{figure}

        Implementing DJ-algorithm on two input qubits requires three qubits including
one work qubit.  The algorithm can be described as
\begin{center}
\begin{equation}
|r \rangle|s \rangle|t \rangle{\stackrel{U_{f}}{\rightarrow}}
|r \oplus f(s,t) \rangle |s \rangle|t \rangle ,
\end{equation}
\end{center}
where $|r \rangle$, $|s \rangle$ and $|t \rangle$ are the states of the work
qubit (I$_{0}$) and of two input (I$_{1}$,I$_{2}$) qubits respectively.  The
eight possible two-bit binary functions are listed in Table 4.

\begin{center}

\bf TABLE 4.\\
The eight possible binary functions ($f_{1}$-$f_{8}$)\\
for the 2-bit DJ algorithm
\normalfont

\vspace{.5cm}

\renewcommand{\tabcolsep}{.3cm}
\begin{tabular}{|c|c|c|c|c|c|c|c|c|c|}
\hline
&
& \multicolumn{2}{c|}{\normalsize CONST.}
& \multicolumn{6}{c|}{\normalsize BAL.}     \\ \cline{3-10}
{\it{s}} & {\it{t}} & {\it{f$_{1}$}} & {\it{f$_{2}$}} & {\it{f$_{3}$}}
& {\it{f$_{4}$}} & {\it{f$_{5}$}} & {\it{f$_{6}$}} & {\it{f$_{7}$}}
& {\it{f$_{8}$}}
\\ \hline
0 & 0 & 0 & 1 & 0 & 1 & 1 & 0 & 1 & 0\\
0 & 1 & 0 & 1 & 0 & 1 & 0 & 1 & 0 & 1\\
1 & 0 & 0 & 1 & 1 & 0 & 1 & 0 & 0 & 1\\
1 & 1 & 0 & 1 & 1 & 0 & 0 & 1 & 1 & 0\\
\hline
\end{tabular}

\end{center}

The pulse scheme 
is same as in Fig.5. Once again,  the transformations corresponding to 
$\it{f}$$_{1}$ and $\it{f}$$_{2}$ are respectively unity operation and a spin 
selective $\pi$$_{x}$ pulse on the work qubit.  The unitary transformations 
encoding the six balanced functions $\it{f}$$_{3}$-$\it{f}$$_{8}$ are implemented 
by selective pulses on the transitions of the work qubit, taken two at a time i.e., 
$[0,0,\pi,\pi]$, $[\pi,\pi,0,0]$, $[\pi,0,\pi,0]$, $[0,\pi,0,\pi]$, $[\pi,0,0,\pi]$, 
and $[0,\pi,\pi,0]$ where 0 denotes no pulse on that particular transition (10,12).  
The results of the algorithm for all the 8 functions are shown in Fig.7.

\begin{figure}
\begin{center}
\hspace{-.5cm}
\psfig{file=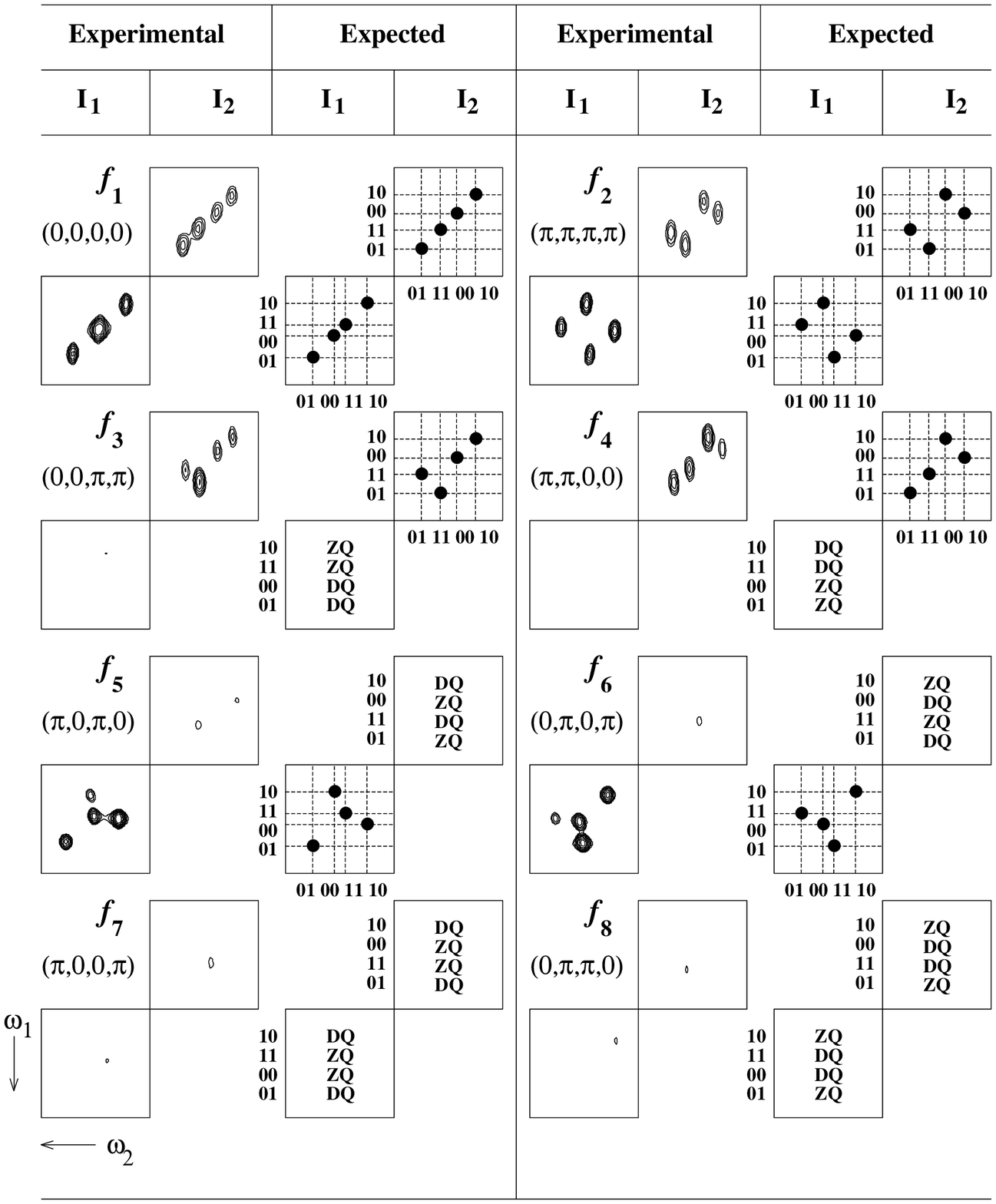,height=12.1cm}
\vspace{-1cm}
\end{center}
\bf{Figure 7.} \normalfont \small The results of DJ
algorithm on 2,3-dibromo-propionic acid (Fig.3a,b) in CDCl$_{3}$
for various functions $f_{1}$-$f_{8}$ listed in Table 4.
I$_{0}$ is the work qubit and I$_{1}$,I$_{2}$ are the input qubits. Only expansions
of the I$_{1}$ and I$_{2}$ parts of the spectra are shown. Expected pattern is
also shown in each case for comparision.  The expected patterns for $f_{2}$ and $f_{4}$
are described in Table 5. 
All experiments were
carried out on a Bruker DRX-500 spectrometer at 300K.  The spin-selective pulses
were 10ms long and the transition-selective pulses were 100-300ms long.  The phase
of the computation pulses were cycled through ($x, -x$) to suppress  the distortions
due to pulse imperfections. 1024 complex data points along t$_{2}$ were acquired for
256 t$_{1}$ values with 2 scans for each t$_{1}$ point.  Zero filling to
1024$\times$1024 complex data points was done prior to the 2D Fourier transformation.
All plots are shown in magnitude mode. \normalsize \\
\end{figure}

 Once again, 
a function is constant only if signals from all the input qubits are present, 
otherwise the function is balanced (Fig.7). Table 5 describes the input-output
correlation for $f_{2}$ and $f_{4}$ in a 2-qubit DJ algorithm.  Here the s- and t- 
transitions are labelled by the states $|r \rangle|t \rangle$ and
$|r \rangle|s \rangle$ respectively.  Other details of Table 5 are similar to 
that of Table 3.

\newpage

\begin{center}
\bf TABLE 5. \\
The input-output correlations for the functions f$_{2}$ and
f$_{4}$ of 2-bit DJ algorithm
\end{center}

\begin{figure}
\vspace{-1.5cm}
\hspace{-2cm}
\psfig{file=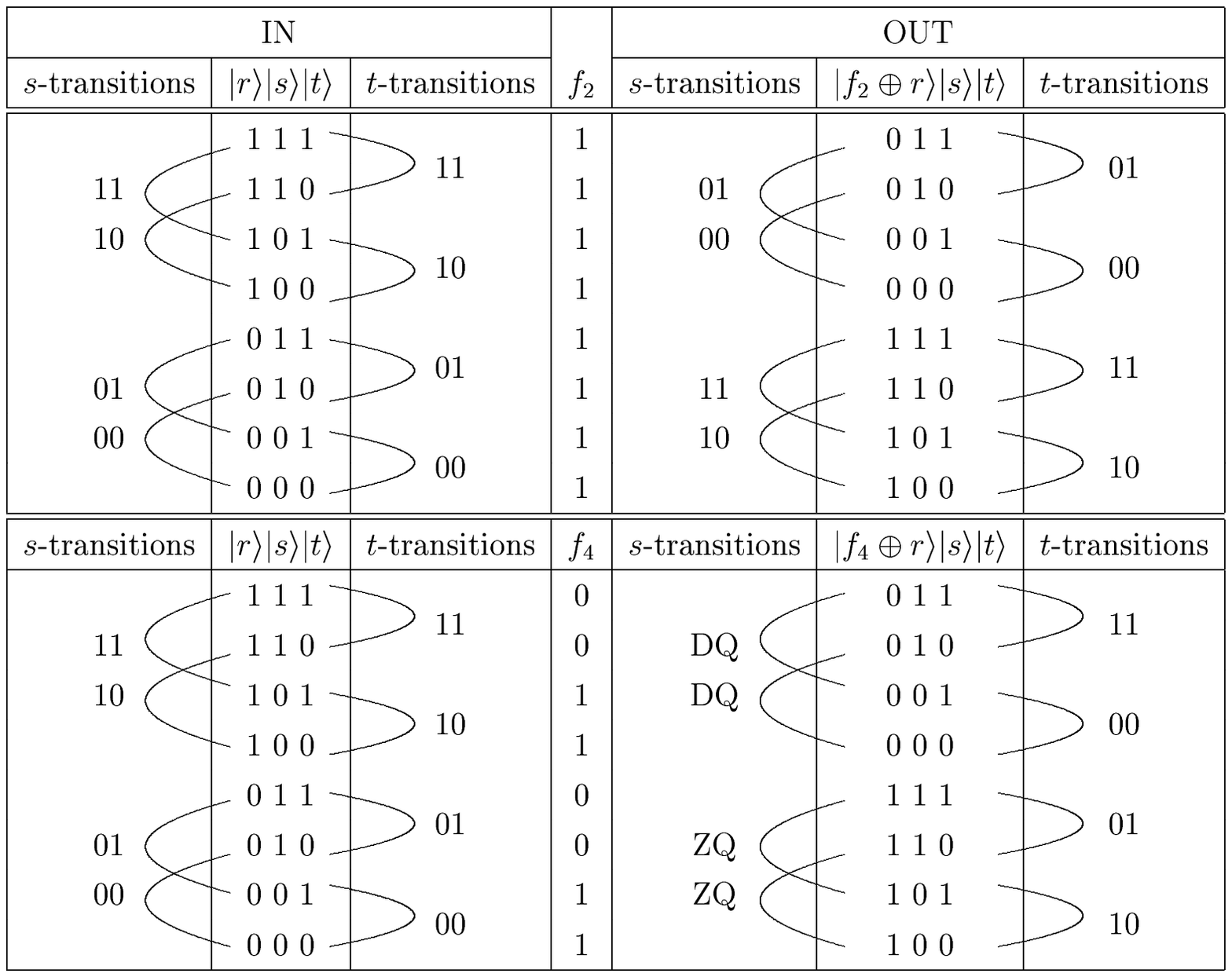,height=16cm}
\vspace{-8cm}
\end{figure}

\section*{Conclusions}
Using two-dimensional NMR quantum computing method, several 2- and 3-qubit gates 
have been implemented on 3 and 4 weakly coupled spin systems and 1- 
and 2-qubit DJ algorithm on 2 and 3 weakly coupled spin systems by 
utilizing spin- and transition-selective pulses.  The use of selective pulses over 
scalar evolution has several advantages as well as some disadvantages. The
advantages of selective pulses are that they lead to simple logic, need only a few 
pulses and work very well.  The disadvantages are that, one needs long low power 
r.f. pulses during which relaxation and r.f. inhomogeneity effects degrade the 
performance of the selective pulses and ideal selectivity may be difficult to 
achieve.  Furthermore, for the selective pulse experiments, one needs resolved 
transitions.  However, many of these difficulties can be overcome by the use of 
heteronuclear spins such as $^{15}$N, $^{13}$C, $^{19}$F, $^{31}$P and 
$^{1}$H, where the magnitudes of the spin-spin couplings are large, the transitions are well 
spread out and selectivity is easier to achieve using pulses of shorter durations.  Further 
improvement can be achieved by orienting molecules in liquid crystal matrices. 
Attempts are continuing in this direction in our as well as in other 
laboratories (18,19).  

\section*{Acknowledgment}
The use of 400 and 500 MHz FTNMR spectrometers of the Sophisticated Instruments
Facility (SIF), Indian Institute of Science, funded by Department of Science
and Technology, New Delhi are gratefully acknowledged.  We also thank Prof.
K. V. Ramanathan and Dr. G. A. Naganagowda of SIF for discussions.

\section*{References}
\begin{itemize}
\item[1.] P. W. Shor,
Polynomial-time algorithms for prime factorization and discrete algorithms
on quantum computer,
SIAM Rev. \bf 41\normalfont, 303-332 (1999).
\item[2.] L. K. Grover, 
Quantum mechanics helps in searching for a needle in a haystack,
Phys. Rev. Lett. \bf 79\normalfont, 325-328 (1997).
\item[3.] D. G. Cory, A. F. Fahmy, and T.F. Havel, 
Ensemble quantum computing by NMR spectroscopy,
Proc. Natl. Acad. Sci. USA \bf 94\normalfont, 1634-1639 (1997).
\item[4.] N. Gershenfeld and I. L. Chuang, 
Bulk spin-resonance quantum computation,
Science \bf 275\normalfont, 350-356 (1997).
\item[5.] I. L. Chuang, L. M. K. Vandersypen, X. Zhou, D. W. Leung and S. Lloyd, 
Experimental realization of a quantum algorithm,
Nature \bf 393\normalfont, 143-146 (1998).
\item[6.] I. L. Chuang, N. Gershenfeld, and M. Kubinec, 
Experimental implementation of fast quantum searching,
Phys. Rev. Lett. \bf80\normalfont, 3408-3411 (1998).
\item[7.] I. L. Chuang, N. Gershenfeld, M. G. Kubinec, and D. W. Leung, 
Bulk quantum computation with nuclear magnetic resonance: theory and experiment,
Proc. Roy. Soc. Lond. A \bf 454\normalfont, 447-467 (1998).
\item[8.] J. A. Jones, R. H. Hansen, and M. Mosca, 
Quantum logic gates and nuclear magnetic resonance pulse sequences,
J. Magn. Reson. \bf135\normalfont, 353-360 (1998).
\item[9.] J. A. Jones and M. Mosca, 
Implementation of a quantum algorithm on a nuclear magnetic resonance quantum computer,
Jl. Chem. Phys. \bf109\normalfont, 1648-1653 (1998).
\item[10.] N. Linden, H. Barjat and R. Freeman, 
An implementation of the Deutsch-Jozsa algorithm on a three-qubit NMR quantum computer,
Chem. Phys. Lett. \bf296\normalfont, 61-67 (1998).
\item[11.] Arvind, Kavita Dorai, and Anil Kumar, 
Quantum entanglement in the NMR implementation of the Deutsch-Jozsa algorithm,
quant-ph/9909067.
\item[12.] Kavita Dorai, Arvind, and Anil Kumar, 
Implementing quantum-logic operations, pseudopure states, and the Deutsch-Jozsa algorithm
using noncommuting selective pulses in NMR,
Phys. Rev. A \bf61\normalfont, 042306/1-7 (2000).
\item[13.] Z. L. Madi, R. Bruschweiler, and R. R. Ernst, 
One- and two-dimensional ensemble quantum computing in spin Liouville space, 
J. Chem. Phys. \bf 109\normalfont, 10603-10611 (1998).
\item[14.] Kavita Dorai and Anil Kumar, 
Cascades of selective pulses on connected single-quantum transitions leading to
the selective excitation of multiple-quantum coherences,
J. Magn. Reson. \bf114\normalfont, 155-162 (1995).
\item[15.] T. Toffoli, Languages and Programming, \it in \normalfont ``Automata"
(J. W. de Bakker and J. van Leeuwen, Eds.), pp. 632-644, Springer (1980).
\item[16.] D. Deutsch and R. Jozsa, 
Rapid solution of problems by quantum computation,
Proc. Roy. Soc. Lond. A \bf439\normalfont, 553-558 (1992).
\item[17.] R. Cleve, A. Ekert, C. Macchiavello, and M. Mosca, 
Quantum algorithms revisited,
Proc. Roy. Soc. Lond. A \bf 454\normalfont, 339-354 (1998).
\item[18.] C. S. Yannoni, M. H. Sherwood, D. C. Miller, I. L. Chuang, L. M. K. Vandersypen,
L. M. K. Vandersypen, and M. G. Kubinec, 
Nuclear magnetic resonance quantum computing using liquid crystal solvents,
Appl. Phys. Lett. \bf 75\normalfont, 3563-3565 (1999).
\item[19.] M. Marjanska, I. L. Chuang, and M. G. Kubinec, 
Demonstration of quantum logic gates in liquid crystal nuclear magnetic resonance,
J. Chem. Phys. \bf 112\normalfont, 5095-5099 (2000).
\end{itemize}

\end{multicols}
\end{document}